\documentclass[12pt]{article}
\usepackage{cite}
\usepackage{enumerate}
\usepackage{ifpdf}

\usepackage{arydshln}
\usepackage{ae}
\usepackage[T1]{fontenc}
\usepackage[ansinew]{inputenc}
\usepackage{csquotes}

\usepackage{amsmath}
\usepackage{relsize}
\usepackage{amssymb}
\usepackage{tikz}
\usepackage{graphicx}
\usepackage{color}
\definecolor{darkblue}{cmyk}{0.9,0.9,0,0}
\definecolor{darkgreen}{rgb}{0,0.55,0}
\usepackage[colorlinks=true,linkcolor=darkblue,citecolor=darkblue,urlcolor=darkblue]{hyperref}
\usepackage{epsfig}
\usepackage{graphicx}
\usepackage[bulletsep]{collref}

\newcommand{\beq}{\begin{equation}}
\newcommand{\eeq}{\end{equation}}
\newcommand{\beqy} {\begin{eqnarray}}
\newcommand{\eeqy} {\end{eqnarray}}
\newcommand{\bsmat}{\begin{smallmatrix}}
\newcommand{\esmat}{\end{smallmatrix}}
\newcommand{\bmat}{\begin{matrix}}
\newcommand{\emat}{\end{matrix}}
\newcommand{\sfrac}[2]{{\textstyle\frac{#1}{#2}}}

\def\({\left(}
\def\){\right)}
\def\[{\left[}
\def\]{\right]}

\def\<{\langle}
\def\>{\rangle}

\def\a{\alpha}

\usepackage{varioref}
\usepackage{makeidx}
\makeindex

\usepackage[english]{babel}
\usepackage{subfig}

\begin{document}

\thispagestyle{empty}

\renewcommand{\thefootnote}{\fnsymbol{footnote}}
\setcounter{page}{1}
\setcounter{footnote}{0}
\setcounter{figure}{0}

\begin{titlepage}

\begin{center}

\vskip 2.3 cm 

\vskip 5mm

{\Large \bf Revisiting $\mathcal{N}=4$ superconformal blocks}
\vskip 15mm

{\large Agnese Bissi and Tomasz {\L}ukowski}

\vspace{1cm}
\centerline{{\it Mathematical Institute, University of Oxford,}}
\centerline{{\it Andrew Wiles Building, Radcliffe Observatory Quarter,}}
\centerline{{\it Woodstock Road, Oxford, OX2 6GG, UK}}
\bigskip
{\tt bissi,lukowski$\bullet$maths.ox.ac.uk}
\end{center}

\vskip 2 cm

\begin{abstract}
We study four-point correlation functions of four generic half-BPS supermultiplets of $\mathcal{N}=4$ SCFT in four dimensions. We use the two-particle Casimir of four-dimensional superconformal algebra to derive superconformal blocks which contribute to the partial wave expansion of such correlators. The derived blocks are defined on analytic superspace and allow us in principle to find any component of the four-point correlator. The lowest component of the result agrees with the superconformal blocks found by Dolan and Osborn. 
\end{abstract}

\bigskip \vfill

\end{titlepage}

\setcounter{page}{1}
\renewcommand{\thefootnote}{\arabic{footnote}}
\setcounter{footnote}{0}

\section{Introduction}
In recent years we observe an increased interest in understanding  Conformal Field Theories (CFT) in dimensions higher than two. One of the most prominent developments comes from the conformal bootstrap approach \cite{Ferrara:1973yt,Ferrara:1973vz,Polyakov:1974gs}. This term stands for a method of constraining the CFT data, namely dimensions and structure constants of primary operators, purely using  symmetries of the model: conformal symmetry, unitarity, the structure of the Operator Product Expansion (OPE) and, most importantly, crossing symmetry of the four-point correlators (or equivalently associativity of the OPE). The latter can be encoded in the so-called bootstrap equation which has already led to numerous numerical predictions for the conformal dimensions \cite{Rattazzi:2008pe} and structure constants \cite{Poland:2010wg,ElShowk:2012hu,Caracciolo:2009bx}, most notably for the three-dimensional Ising model \cite{ElShowk:2012ht,El-Showk:2014dwa}. 
This approach has been extended also to supersymmetric conformal theories, in various space-time dimensions in \cite{Beem:2015aoa,Bobev:2015vsa,Beem:2013qxa,Alday:2013opa,Alday:2014qfa,Chester:2014fya,Bobev:2015jxa,Beem:2014zpa,Khandker:2014mpa,Fitzpatrick:2014oza,Poland:2010wg,Vichi:2011ux,Poland:2011ey,Bashkirov:2013vya,Berkooz:2014yda}.
In this paper we focus on the case of maximally supersymmetric conformal field theories in four dimensions -- $\mathcal{N}=4$ SCFT. The prime example of such theory is $\mathcal{N}=4$ super-Yang-Mills (SYM), which attracted a lot of interest in recent years after it emerged as the first example of the AdS/CFT duality \cite{Maldacena:1997re}. 

One of the reasons CFTs attract so much attention is the fact that conformal symmetry significantly simplifies the form of correlators in such theories. In particular, it completely fixes the space-time dependence of two- and three-point correlators and constrains all higher-point functions to depend non-trivially only on conformal cross-ratios. In $\mathcal{N}=4$ SCFT conformal symmetry is enhanced to a larger superconformal symmetry. Such enhancement helps to further constrain the structure of correlators. The implications of supersymmetry are most studied for operators belonging to short supermultiplets -- supermultiplets satisfying BPS shortening conditions. One of those are the so-called "chiral" primary, half-BPS operators which are of most interest for this paper. For these operators superconformal symmetry leads to various non-renormalizability properties \cite{Intriligator:1998ig,Intriligator:1999ff,Eden:1999gh,Petkou:1999fv,Howe:1999hz,Heslop:2001gp}. This can be traced back to the fact that these supermultiplets are shorter than generic ones and as such depend only on half of the fermionic coordinates of $\mathcal{N}=4$ superspace (8 instead of 16). It can be shown that all correlators which have less than 32 fermionic degrees of freedom -- number of supercharges for $\mathcal{N}=4$ superalgebra -- are completely fixed by supersymmetry and they do not acquire any quantum correction. This is the case for two- and three-point correlators of half-BPS supermultiplets. It follows that for all operators in these supermultiplets the conformal dimensions as well as structure constants are independent of the coupling constant and they are fixed completely by their classical contributions. In the case when the number of fermionic degrees of freedom equals the number of supercharges, we get partial non-renormalizability. Such correlators will depend only on one non-trivial function of the coupling constant. This is the case of four-point correlators of four half-BPS supermultiplets and also for three-point correlators of two half-BPS and one generic supermultiplet \cite{Eden:2001ec} or two-point correlators of generic multiplets. The latter implies that for a generic supermultiplet the conformal dimensions of all descendants are related to the conformal dimension of parent superconformal primary.

One of the main features of any CFT is the fact that its OPE expansion simplifies. The space-time structure of the OPE is completely fixed by the conformal symmetry and structure constants for the descendants are fixed by structure constants of their primaries. It allows in particular to expand four-point correlators in conformal partial waves, after taking a double OPE expansion, as a weighted sum of conformal blocks over all multiplets present in the OPE of external operators. Conformal blocks resum contributions coming from a complete multiplet from the OPE expansions. In the presence of supersymmetry it is a natural question to ask whether it is possible to write four-point functions as a sum of superconformal blocks. This is not always possible since for generic operators the three-point functions of superdescendants are not fixed by their superconformal primaries (as opposed to their dimensions). As we mentioned before it is, however, true when two of the operators belong to a short supermultiplet. It is possible then to define superconformal blocks at least in the case of the four-point function of four half-BPS supermultiplets. This is indeed the goal of this paper. The study of the relation between conformal blocks and supersymmetry was started in the series of papers by Dolan and Osborn \cite{Dolan:2001tt,Dolan:2004iy} \footnote{The same results for four dimensional $\mathcal{N}$=4 SYM have been obtained also in \cite{Dolan:2004mu}.}. Their work was based  on previous results on superconformal Ward identities, see \cite{Nirschl:2004pa}. As a result of their analysis, Dolan and Osborn defined a function they called a superconformal block which enters in the partial wave decomposition of the lowest component of the four-point correlator of four half-BPS superfields. In this paper we call the superconformal block a function defined on the full superspace, depending also on the fermionic degrees of freedom. Note that this terminology differs from the one of Dolan and Osborn who focused on the lowest component of superconformal block defined here. We will construct such a function before the end of this paper. Then, any four-point correlator of half-BPS supermultiplets can be decomposed as a weighted sum of such superconformal blocks, which encode the contribution of superconformal primaries and all their superconformal descendants. 

The closed form of ordinary conformal blocks is known in four dimensions \cite{Dolan:2000ut}. In their work on superconformal blocks, Dolan and Osborn used superconformal Ward identities to relate various conformal blocks for operators belonging to the same superconformal multiplet. Afterwards, they were able to sum all such contributions and obtain a closed expression for the full supermultiplet. There exists, however, an  alternative approach to finding superconformal block which will be used in this paper. As for the ordinary conformal blocks, it is possible to find their form using the fact that they are eigenvectors of the two-particle quadratic Casimir, see e.~g.~\cite{Bobev:2015jxa,Fitzpatrick:2014oza}. In this paper we construct the quadratic Casimir for the $\mathfrak{psu}(2,2|4)$ algebra -- the $\mathcal{N}=4$ superconformal algebra in four dimensions. It will provide us with a differential equation, defined on the analytic superspace, which we subsequently solve and give an explicit form of the superconformal blocks. It turns out that there are two solutions of this equation, one corresponding to the long representations and the other to the short representations. When we evaluate the lowest component of these blocks we recover formulas found in \cite{Dolan:2004iy}.

The paper is organized as follows: In Section \ref{Sec:General} we introduce our notations and explain the general strategy of the paper. In Section \ref{Sec:Analytic} we define the analytic superspace which will provide us with a proper set of variables to describe superconformal blocks. We also write the explicit form of the Casimir represented on that space. Section \ref{Sec:OPE} focuses on the OPE of two half-BPS supermultiplets
 and is followed by the discussion on superconformal Ward identities in Section \ref{Sec:Ward}. After gathering all ingredients of our construction, we proceed in Section \ref{Sec:Solution} with the explicit form of the Casimir differential equation and solutions of it. We end with conclusions and outlook. The main text is followed by two appendices with some technical details of the construction.

\section{General framework}\label{Sec:General}
\subsection{Definition of operators}
In this paper we focus on the half-BPS states in $\mathcal{N}=4$ SCFT. The most natural setting to describe these operators and their correlators is the analytic superspace \cite{Howe:1995md,Hartwell:1994rp}. To define it, in addition to the Minkowski space parametrized by $x^{\alpha\dot\alpha}$, one introduces an auxiliary bosonic space with coordinates $y^{aa'}$, that parametrize the $SU(4)$ R-symmetry. This bosonic space is extended to full analytic superspace by including eight Grassmann-odd coordinates which we denote by $\lambda^{\alpha a'}$ and $\pi^{a\dot\alpha}$. All indices $\alpha,\dot\alpha,a,a'$ are two-component spinor indices, see App.~\ref{App:Conventions} for our conventions.

The superfields are naturally expressed in the analytic superspace as
\begin{align}\label{superfield}
\mathcal{T}^{(p)}(x,y,\lambda,\pi)&=\exp(\lambda^{\alpha a'}Q_{\alpha a'}+\pi^{a\dot\alpha}Q_{a\dot\alpha})O^{(p)}(x,y)\\\label{superfield.explicit}&=O^{(p)}(x,y)+\lambda^{\alpha a'}\Psi^{(p)}_{\alpha a'}(x,y)+\pi^{a\dot\alpha}\bar\Psi^{(p)}_{a\dot\alpha}(x,y)+\ldots\,,
\end{align}
where the lowest components are single-trace, half-BPS scalar operators $O^{(p)}$ built out completely of the six elementary fields $\phi_{AB}=-\phi_{BA}$ with $A,B=1,\ldots,4$ and belonging to the $[0,p,0]$ representation of the $SU(4)$ R-symmetry group. In order to keep track of the R-symmetry indices we introduce auxiliary Y-variables. Explicitly, the fields take the form
\begin{equation}
O^{(p)}(x,y)=O_{A_1 B_1,\ldots,A_p B_p} Y^{A_1 B_1}\ldots Y^{A_p B_p}\,,
\end{equation}
where $Y^{AB}$ are related to the analytic space coordinates as
\begin{equation}
Y^{AB}=\left(\begin{tabular}{cc}$\epsilon^{ab}$&$-y^{ab'}$\\$y^{ba'}$&$\epsilon^{a'b'}y^2$\end{tabular}\right).
\end{equation}
The operators $O_{A_1 B_1,\ldots,A_p B_p}$ are symmetric traceless rank $p$ tensors formed out of gauge invariant traces of
the elementary scalar fields $\phi_{AB}$. The remaining terms in the expansion \eqref{superfield.explicit} are superconformal descendants of $O^{(p)}(x,y)$ and for future reference we write explicitly just the first two fermionic components $\Psi$ and $\bar\Psi$.
 Of most interest is the case when $p=2$, since in that case the superfield contains as its components both conserved currents of the theory and the stress-energy tensor. The superfield \eqref{superfield} satisfies a half-BPS shortening condition and as a consequence it depends only on 4 chiral and 4 antichiral Grassmann-odd variables, instead of the 16 Grassmann-odd parameters of the full superspace.  

\subsection{Correlation functions of half-BPS states}
Having defined the operators of interest we proceed to study their correlation functions. Similar to the standard conformal case, the symmetry of the model restricts the form of such correlators significantly. It turns out that superconformal symmetry is sufficient to completely fix the form of the two-point and three-point correlation functions of $\mathcal{T}^{(p)}$, namely, they are protected from quantum corrections \cite{Lee:1998bxa,D'Hoker:1998tz,Howe:1998zi,Penati:1999ba,Penati:2000zv}. The first non-trivial unprotected quantity is the four-point correlator
\begin{equation}\label{4point}
\mathcal{A}^{\{ p_i\}}=\langle \mathcal{T}^{(p_1)}(x_1,y_1,\lambda_1,\pi_1)\mathcal{T}^{(p_2)}(x_2,y_2,\lambda_2,\pi_2)\mathcal{T}^{(p_3)}(x_3,y_3,\lambda_3,\pi_3)\mathcal{T}^{(p_4)}(x_4,y_4,\lambda_4,\pi_4)\rangle\,,
\end{equation}
with $p_1\leq p_2\leq p_3\leq p_4$, which will be of main interest in this paper. For brevity we have introduced the notation $\{ p_i\}=(p_1,p_2,p_3,p_4)$. The superconformal symmetry restricts the form of four-point correlator, which in turn depends only on its lowest component given by the correlator of the scalar fields \cite{Korchemsky:2015ssa}
 \begin{equation}\label{4pointlowest}
 A^{\{ p_i\}}(x_i,y_i)=\mathcal{A}^{\{ p_i\}}|_{\lambda_i=\pi_i=0}=\langle O^{(p_1)}(x_1,y_1)O^{(p_2)}(x_2,y_2)O^{(p_3)}(x_3,y_3)O^{(p_4)}(x_4,y_4)\rangle\,.
 \end{equation}
Four-point correlators of all other components of the supermultiplet are related to \eqref{4pointlowest} by superconformal Ward identities \cite{Belitsky:2014zha,Korchemsky:2015ssa}.

\subsection{Partial wave decomposition}
One of the main features of quantum field theories is the notion of Operator Product Expansion. For a generic CFT the coefficients appearing in the OPE are related to the three-point structure constants and as such allow to fix any four-point correlator using only information of lower-point functions. Moreover, in $\mathcal{N}=4$ SCFT, the three-point function of two half-BPS operators with any non-protected operator is related to three-point function of these half-BPS operators with the superconformal primary of the latter \cite{Eden:2001ec}. Hence superconformal symmetry  allows us to organize the expansion as a decomposition in superconformal blocks
\begin{equation}
\mathcal{A}^{\{ p_i\}}=\frac{y_{12}^{p_1+p_2}y_{34}^{p_3+p_4}}{x_{12}^{p_1+p_2}x_{34}^{p_3+p_4}}\left(\frac{x_{24} y_{14}}{x_{14} y_{24}} \right)^{p_{12}} \left(\frac{x_{14} y_{13}}{x_{13} y_{14}} \right)^{p_{34}}\sum_{\Delta,\ell,\mathcal{R}}c_{\Delta,\ell,\mathcal{R}}^{p_1,p_2}c_{\Delta,\ell,\mathcal{R}}^{p_3,p_4} \,\mathfrak{g}^{\{p_i\}}_{\Delta,\ell,\mathcal{R}}(x_i,y_i,\lambda_i,\pi_i)\,,
\end{equation}
with $p_{ij}=p_i-p_j$ and where the sums runs over all superconformal primaries appearing in the OPE of $O^{(p_1)}\times O^{(p_2)}$ as well as $O^{(p_3)}\times O^{(p_4)}$, labeled by their dimension $\Delta$, spin $\ell$ and the R-symmetry representation $\mathcal{R}$. The coefficients $c_{\Delta,\ell,\mathcal{R}}^{p_i,p_j}$ are properly normalized three-point functions of $O^{(p_i)}$ and $O^{(p_j)}$ with the intermediate superconformal primary $O_{\Delta,\ell,\mathcal{R}}$. The superconformal blocks $\mathfrak{g}^{\{p_i\}}_{\Delta,\ell,\mathcal{R}}(x_i,y_i,\lambda_i,\pi_i)$ depend on the dimensions of external operators $\{ p_i\}$ as well as the dimension $\Delta$, spin $\ell$ and the R-symmetry representation $\mathcal{R}$ of the intermediate states. Notice that the superconformal symmetry implies that the lowest component of the superconformal block
\begin{equation}\label{lowest.block}
g^{\{p_i\}}_{\Delta,\ell,\mathcal{R}}(u,v,\sigma,\tau)=\mathfrak{g}^{\{p_i\}}_{\Delta,\ell,\mathcal{R}}(x_i,y_i,\lambda_i,\pi_i)|_{\lambda_i=\pi_i=0}
\end{equation}
 is a function of the conformal cross-ratios
\begin{equation}\label{conf.cross}
u=\frac{x_{12}^2 x_{34}^2}{x_{13}^2 x_{24}^2}=z\, \bar z\,,\qquad v=\frac{x_{23}^2x_{14}^2}{x_{13}^2x_{24}^2}=(1-z)(1-\bar z)\,,
\end{equation}  
as well as the harmonic cross-ratios\footnote{We would like to bring the reader attention to the fact that our conventions differ from the standard definition one can find in the literature, see e.~g.~\cite{Nirschl:2004pa}} 
\begin{equation}\label{harm.cross}
\sigma=\frac{y_{12}^2 y_{34}^2}{y_{13}^2 y_{24}^2}=\alpha \,\bar\alpha\,,\qquad \tau=\frac{y_{23}^2y_{14}^2}{y_{13}^2y_{24}^2}=(1-\alpha)(1-\bar\alpha)\,.
\end{equation}
Here, we introduced a convenient parametrization of cross-ratios with the use of $z,\bar z,\alpha,\bar\alpha$, which turns out to be a natural set of variables to describe superconformal blocks.

\subsection{Superconformal Casimir}
The aim of this paper is to find an explicit form of the superconformal blocks $\mathfrak{g}^{\{p_i\}}_{\Delta,\ell,\mathcal{R}}$. These blocks are the eigenfunctions of the two-particle quadratic Casimir $\mathcal{C}_{12}$ of the maximal superconformal algebra in four dimension, namely, $\mathfrak{psu}(2,2|4)$
\begin{equation}\label{Casimireigenproblem}
\mathcal{C}_{12} \,\mathfrak{G}^{\{p_i\}}_{\Delta,\ell,\mathcal{R}}(x_i,y_i,\lambda_i,\pi_i)=\kappa_{\Delta,\ell,\mathcal{R}}\,\mathfrak{G}^{\{p_i\}}_{\Delta,\ell,\mathcal{R}}(x_i,y_i,\lambda_i,\pi_i)\,,
\end{equation}
with
\begin{equation}\label{capitalG}
\mathfrak{G}^{\{p_i\}}_{\Delta,\ell,\mathcal{R}}(x_i,y_i,\lambda_i,\pi_i)=\left(\frac{x_{24} y_{14}}{x_{14} y_{24}} \right)^{p_{12}} \left(\frac{x_{14} y_{13}}{x_{13} y_{14}} \right)^{p_{34}}\mathfrak{g}^{\{p_i\}}_{\Delta,\ell,\mathcal{R}}(x_i,y_i,\lambda_i,\pi_i)\,.
\end{equation}
Here and in the following we use capital letters, e.~g.~$\mathfrak{G}$, to distinguish expressions containing the inhomogeneous prefactor as in \eqref{capitalG}, while lower case letters,  such as $\mathfrak{g}$, denote functions invariant under the action of dilatation. We call both objects superconformal blocks. 

 Denoting by $J_{i,A}^B$ the generators of the $\mathfrak{psu}(2,2|4)$ algebra acting on the operator at position $i$, the schematic form of the two-particle Casimir is
\begin{equation}
\mathcal{C}_{12}=\sum_{A,B}\left(J_{1,A}^{B}+J_{2,A}^{B}\right)\left(J_{1,B}^{A}+J_{2,B}^{A}\right)\,.
\end{equation}
In the following we  give an explicit form of the generators and the Casimir itself realized as differential operators acting on the analytic space. In that case the eigenproblem \eqref{Casimireigenproblem} reduces to solving a second order differential equation.

\subsection{Superconformal Ward identities}
As we pointed out already, it is enough to specify the four-point correlator of the lowest component in the supermultiplet and use the superconformal Ward identities to recover the full supermultiplet. The same statement holds true for the superconformal block. In the following we will focus on the scalar block \eqref{lowest.block} corresponding to the decomposition of the lowest component four-point function $A^{\{ p_i\}}$ as in \eqref{4pointlowest}. However, even in that case our starting point is the full superconformal block equation \eqref{Casimireigenproblem}, since the action of the Casimir operator mixes various components of the multiplet. As we will see soon the only relevant elements of the expansion of the superconformal block are  
\begin{align}\label{superblock.expansion}
\mathfrak{G}^{\{p_i\}}_{\Delta,\ell,\mathcal{R}}(x_i,y_i,\lambda_i,\pi_i)=G^{\{p_i\}}_{\Delta,\ell,\mathcal{R}}(x_i,y_i)+\lambda_{1}\,\pi_2\, Q^{\{p_i\}}_{\Delta,\ell,\mathcal{R}}(x_i,y_i)
+\pi_{1}\,\lambda_2\, \bar{Q}^{\{p_i\}}_{\Delta,\ell,\mathcal{R}}(x_i,y_i)+\ldots\,,
\end{align} 
where $\ldots$ stands for the contributions which decouple after substituting into the equation \eqref{Casimireigenproblem} and projecting to its lowest component. Notice that we suppressed all the indices in \eqref{superblock.expansion} and following formulas, however, it is straightforward to reintroduce them. Here, the function $G^{\{p_i\}}_{\Delta,\ell,\mathcal{R}}$ is the block entering in the decomposition of the lowest four-point function
\begin{equation}
 A^{\{p_i\}}(x_i,y_i)=\frac{y_{12}^{p_1+p_2}y_{34}^{p_3+p_4}}{x_{12}^{p_1+p_2}x_{34}^{p_3+p_4}} \sum_{\Delta,\ell,\mathcal{R}}c_{\Delta,\ell,\mathcal{R}}^{p_1,p_2}c_{\Delta,\ell,\mathcal{R}}^{p_3,p_4} \,G^{\{p_i\}}_{\Delta,\ell,\mathcal{R}}(x_i,y_i)\,,
\end{equation} 
while the functions $Q^{\{p_i\}}_{\Delta,\ell,\mathcal{R}}(x_i,y_i)$ and $\bar{Q}^{\{p_i\}}_{\Delta,\ell,\mathcal{R}}(x_i,y_i)$ enter into an analogous decomposition of $\langle \Psi^{(p_1)}\bar\Psi^{(p_2)}O^{(p_3)}O^{(p_4)}\rangle$ and $\langle \bar\Psi^{(p_1)}\Psi^{(p_2)}O^{(p_3)}O^{(p_4)}\rangle$, respectively.

In order to find the form of the superconformal block we disentangle the lowest component and write an explicit equation for  $G^{\{p_i\}}_{\Delta,\ell,\mathcal{R}}$. To do this we use superconformal Ward identities which relate both $Q^{\{p_i\}}_{\Delta,\ell,\mathcal{R}}$ and $\bar Q^{\{p_i\}}_{\Delta,\ell,\mathcal{R}}$ to $G^{\{p_i\}}_{\Delta,\ell,\mathcal{R}}$. In the analytic space the solution to Ward identities was written explicitly in \cite{Belitsky:2014zha}.
Combining the superconformal Casimir equation and the superconformal Ward identities amounts of having a second order differential equation for the scalar block $G^{\{p_i\}}_{\Delta,\ell,\mathcal{R}}$.

Moreover, the possible structure of the four-point function of the lowest component is determined also by superconformal Ward identities \cite{Nirschl:2004pa}.  We define a function of the cross-ratios $\hat{A}^{\{p_i\}}(u,v,\sigma, \tau)$ in the following way
\begin{equation}
A^{\{ p_i\}}(x_i,y_i)=\frac{y_{12}^{p_1+p_2}y_{34}^{p_3+p_4}}{x_{12}^{p_1+p_2}x_{34}^{p_3+p_4}}\left(\frac{x_{24} y_{14}}{x_{14} y_{24}} \right)^{p_{12}} \left(\frac{x_{14} y_{13}}{x_{13} y_{14}} \right)^{p_{34}} \hat{A}^{\{p_i\}}(u,v,\sigma, \tau)\,.
\end{equation}
Now, we can use superconformal Ward identities, together with the symmetries of the four-point function under the exchange of $z$ and $\bar{z}$ as well as $\alpha$ and $\bar{\alpha}$ and the fact that the four-point function depends polynomially on the R-symmetry variables, to fix the structure of $\hat{A}$ to be
\begin{align}\nonumber
\hat{A}^{\{ p_i\}}(z,\bar{z},\alpha, \bar{\alpha})&=\left(\frac{\alpha\bar\alpha}{z \bar z}\right)^{\sfrac{p_{34}}{2}} k +\left(\frac{\alpha\bar\alpha}{z \bar z}\right)^{\sfrac{p_{34}}{2}}\frac{ (z-\alpha ) (z-\bar\alpha) (\bar z-\alpha ) (\bar z -\bar\alpha)}{ (\alpha -\bar\alpha) (z-\bar z)}\times\\\nonumber
&\times\left(-\frac{ f(z,\alpha )}{\alpha\,  z\, (\bar z -\bar\alpha) }+\frac{  f(z,\bar\alpha)}{\bar\alpha\, z\, (\bar z -\alpha)} +\frac{  f(\bar z,\alpha )}{\alpha  \,\bar z\, (z-\bar\alpha)}-\frac{f(\bar z,\bar\alpha)}{\bar \alpha\, \bar z\,(z-\alpha)}\right)\\ \label{wardlowest}
&+(z-\alpha)(z-\bar\alpha)(\bar z-\alpha)(\bar z-\bar\alpha)F(z,\bar{z},\alpha,\bar{\alpha})\,,
\end{align}
where the functions $F(z,\bar{z},\alpha,\bar{\alpha})$, $f(z,\alpha)$ and $k$ are arbitrary.
In section \ref{Sec:solutionToCasimir} we will see how to interpret the various contributions  in \eqref{wardlowest}.  

\section{Superconformal algebra in analytic superspace}\label{Sec:Analytic}

The maximal superconformal algebra in four dimensions is $\mathfrak{psu}(2,2|4)$. It is composed of the bosonic Poincar\'e and R-symmetry subalgebras, together with 32 supercharges. For the half-BPS operators half of the supercharges annihilates the operators. This leads to significant simplifications compare with generic, non-protected operators and allows to restrict the full $\mathcal{N}=4$ superspace to its analytic subspace defined in \cite{Howe:1995md,Hartwell:1994rp}. As we already pointed out, the analytic superspace consists of the Minkowski space together with an auxiliary space, locally identical to the Minkowski space, which parametrizes the $SU(4)$ R-symmetry of the model. This bosonic space is then supplemented by 8 Grassmann-odd variables $\lambda^{\alpha a'}$ and $\pi^{a\dot\alpha}$, half of the full superspace coordinates. Here, we split the $SU(4)$ index $A=(\alpha,\dot\alpha)$, this however does not break the $SU(4)$ R-symmetry. Analytic superspace coordinates can be nicely combined in the $(2|2)\times (2|2)$ supermatrix
\begin{equation}
\left( \begin{array}{c:c}x^{\alpha\,\dot\alpha}&\lambda^{\alpha \, a'}\\\hdashline
\pi^{a\,\dot\alpha}&y^{a\,a'}\end{array}\right)\,,
\end{equation}
where all indices can take just two values: $\alpha=1,2$, $\dot\alpha=\dot{1},\dot{2}$, $a=1,2$, $a'=1',2'$. All superfields are now defined as functions of these coordinates, more precisely, as polynomials in fermionic and $y$ variables, and a generic function on the Minkowski space.

The next step in our construction is to describe how the $\mathfrak{psu}(2,2|4)$ superconformal algebra is realized on the analytic superspace \cite{Howe:1999hz}. We write the explicit form for all generators in the appendix \ref{App:Generators}. They can be organized as the $(4|4)\times (4|4)$-supermatrix, all elements being differential operators acting on the analytic superspace, in the following way 
\begin{equation}\label{generators.matrix}
\left( \begin{tabular}{c:c|c:c}$L_{\alpha}^{\,\,\,\beta}$ & $P_{\alpha\dot\beta}$ & $Q^{\,\,\,b}_{\alpha}$ & $Q_{\alpha b'}$\\
\hdashline
$K^{\dot\alpha\beta}$ & $\bar{L}^{\dot\alpha}_{\,\,\,\dot\beta}$ & $S^{\dot\alpha\, b}$ & $S_{\,\,\,b'}^{\dot\alpha}$\\
\hline
$S^{\,\,\,\beta}_{a}$ & $Q_{a\dot\beta}$ & $R_{a}^{\,\,\,b}$ & $P'_{a\,b'}$\\
\hdashline
$S^{a'\beta}$ & $Q^{a'}_{\,\,\,\dot\beta}$ & $K'^{a'b}$ & $\bar{R}^{a'}_{\,\,\,b'}$ \end{tabular}\right).
\end{equation}
This has to be supplemented by the dilatation operator $D$ and its R-symmetry counterpart $D'$. The upper-left part of \eqref{generators.matrix} is the Poincar\'e algebra while the lower-right corner describe R-symmetry.   

All operators in $\mathcal{N}=4$ SCFT are organized by the superconformal symmetry into supermultiplets which are highest weight representations of the $\mathfrak{psu}(2,2|4)$ algebra. They consist of a unique superconformal primary operator which is annihilated by all positive roots of the algebra, namely, by all generators:
\begin{equation}\label{positive.roots}
\{S_{\,\,\,b'}^{\dot\alpha},S^{\,\,\,\beta}_{a},S^{\dot\alpha b},S^{a'\beta} ,K^{\dot\alpha\beta}, K'^{a'b},L_{\alpha}^{\,\,\,\beta}(\alpha>\beta),\bar{L}^{\dot\alpha}_{\,\,\,\dot\beta}(\dot\alpha>\dot\beta),R_{a}^{\,\,\,b}(a>b),\bar{R}^{a'}_{\,\,\,b'}(a'>b')  \}\,.
\end{equation} 
The action of the diagonal generators in \eqref{generators.matrix} on the superconformal primary specifies representation labels, while the negative roots create new operators (descendants) belonging to the same highest-weight module. All representations of the superconformal algebra are parametrized by two $SU(2)$ spins $j_1,j_2$, conformal dimension $\Delta$ as well as R-symmetry representation labels $[r_1,q,r_2]$. In our derivation we distinguish two types of supermultiplets. Firstly, we have external half-BPS operators which take the form \eqref{superfield}. These are protected operators, i.e.~they do not acquire any anomalous dimension, which are annihilated by half of the lowering supercharges, in addition to positive roots \eqref{positive.roots}. Secondly, there are the operators which appear in the OPE decomposition of external fields that can be either protected (half- and quarter-BPS) or non-protected. In our discussion all relevant operators have $j_1=j_2\equiv\ell$, since in the OPE of two supermultiplets \eqref{superfield} there are only completely symmetric traceless tensors of rank $\ell$. For the external operators the R-symmetry labels are given by $[0,p,0]$. In the intermediate channel we can have, however, more general operators transforming in a $[r,q,r]$ R-symmetry representation.

Superconformal blocks are eigenvectors of the quadratic Casimir of the $\mathfrak{psu}(2,2|4)$ algebra. The latter is an element of the universal enveloping algebra, quadratic in generators and commutes with all of them and was introduced for the $\mathfrak{psu}(2,2|4)$ algebra in \cite{PROP:PROP2190350705}. With our notation it takes the following form, see e.~g.~\cite{Beisert:2004ry},
\begin{align}
\mathcal{C}_{12}&=L_{\alpha}^{\,\,\, \beta}L_{\beta}^{\,\,\,\alpha}+\bar{L}^{\dot\alpha}_{\,\,\,\dot\beta}\bar{L}^{\dot\beta}_{\,\,\,\dot\alpha}-\{ P_{\alpha\dot\beta},K^{\dot\beta\alpha}\}+D^2-R_{a}^{\,\,\,b}R_{b}^{\,\,\,a}-\bar{R}_{\,\,\,b'}^{a'}\bar{R}_{\,\,\,a'}^{b'}-\bar{D}^2+\{ P'_{ab'},K'^{\,b' a} \}\\
&+[Q_{\alpha}^{\,\,\, b},S_{b}^{\,\,\,\alpha}]+[Q_{\alpha b'},S^{b'\alpha}]-[S^{\dot\alpha b},Q_{b\dot\alpha}]-[S^{\dot\alpha}_{\,\,\,b'},Q^{b'}_{\dot\alpha}]\,.
\end{align}
In order to write down the eigenproblem equation for superconformal blocks \eqref{Casimireigenproblem} we need also to know the eigenvalue of quadratic Casimir $\kappa_{\Delta,\ell,[r_1,q,r_2]}$ for a given supermultiplet with dimension $\Delta$, spin $\ell$ and in the R-symmetry representation $\mathcal{R}=[r_1,q,r_2]$. It is given by
\begin{equation}\label{Casimir.eigen}
\kappa_{\Delta,\ell,[r_1,q,r_2]}=(\Delta+4)\Delta+\ell(\ell+2)-\sfrac{1}{2}r_1(r_1+2)-\sfrac{1}{2}r_2(r_2+2)-\sfrac{1}{4} (2 q + r_1 + r_2)^2 - 2(2 q + r_1 + r_2)\,.
\end{equation}

\section{Structure of the OPE}\label{Sec:OPE}
In this paper we study the four-point function of half-BPS scalar operators of any protected dimension $p_1$, $p_2$, $p_3$ and $p_4$. Using the OPE, we can decompose this four-point function into all possible $SU(4)$ R-symmetry representations appearing both in the tensor product $[0,p_1,0] \times [0,p_2,0] $ and  $[0,p_3,0] \times [0,p_4,0] $. The tensor product of two R-symmetry representations takes the following form \cite{Dolan:2002zh}
\begin{equation} \label{su4reps}
[0,p_1,0] \times [0,p_2,0] = \sum_{r=0}^{p_1} \sum_{s=0}^{p_1-r}[r,p_2-p_1+2 s,r] \,,
\end{equation}
for $p_2 \geq p_1$. The operators appearing in the product \eqref{su4reps} in general belong to different supermultiplet \cite{Nirschl:2004pa}. Schematically, the OPE of two half-BPS operators organizes as \footnote{Notice that we are interested in unitary representations, therefore we are left only with short and semi-short multiplet of type $\mathcal{B}$ and $\mathcal{C}$. For a detailed discussion on this point see \cite{Dolan:2002zh}.}
\begin{align} \nonumber
\mathcal{B}_{[0,p_1,0]} \times \mathcal{B}_{[0,p_2,0]}&=\sum_{\substack{0\leq m \leq n \leq p_1\\ \ell=0}} \mathcal{B}_{[n-m,p_2-p_1+2m,n-m]} + \sum_{\substack{0\leq m \leq n \leq p_1-1\\ \ell \geq 0}} \mathcal{C}_{[n-m,p_2-p_1+2m,n-m],\ell} \\\label{OPE}
&+\sum_{\substack{0\leq m \leq n \leq p_1-2\\ \ell \geq 0}} \mathcal{A}_{[n-m,p_2-p_1+2m,n-m],\Delta,\ell}\,.
\end{align}
where various supermultiplets on the right hand side of \eqref{OPE} match elements of the full classification of unitary irreducible representations of $\mathfrak{psu}(2,2|4)$ introduced in \cite{DOBREV1985127}:
\begin{itemize}
\item The supermultiplet $\mathcal{B}_{[r,q,r]}$ is called short and it contains half-BPS operators if $r=0$ and quarter-BPS operators if $r>0$, all with spin $\ell=0$. The dimension of the corresponding highest-weight state is fixed and depends on the representation, more precisely
\begin{align}
[0,q,0]: &\quad \Delta=q\,,  &\text{$\sfrac{1}{2}$-BPS}\,,\\\label{short.dim}
[r,q,r]:&\quad \Delta=q+2r\,,  &\text{$\sfrac{1}{4}$-BPS}\,.
\end{align}
\item The supermultiplet $\mathcal{C}_{[r,q,r],\ell}$ is called semi-short and it contains current-like operators, which can be half-BPS or quarter-BPS. The dimension is also protected from quantum correction and it is given by
 \begin{equation}\label{semishort.dim}
 \Delta= \ell+2+2 r+q\,.
 \end{equation}
\item The supermultiplet $\mathcal{A}_{[r,q,r],\Delta,\ell}$ is called long. The operators belonging to this supermultiplet generically acquire an anomalous dimension. However, unitarity implies that 
 \begin{equation}
 \Delta >  \ell+2+2 r+q\,.
 \end{equation}
 \end{itemize}
Notice that in general it is possible to consider any particular semi-short multiplets as a part of long multiplet at the expense of introducing another semi-short multiplet \cite{Dolan:2002zh}. This generates ambiguities in defining the conformal partial wave expansion in free field theories. However, this fact will not influence our further discussion.

\section{Superconformal Ward identities}\label{Sec:Ward}
The explicit form of the $\mathcal{N}=4$ SYM half-BPS supermultiplet is given in \eqref{superfield} where each component of the superfield can be expressed as a differential operator acting on the lowest component of the superfield. The specific form of this differential operator is given by solving the superconformal Ward identities. The same statement is true for correlation functions involving any component of the supermultiplet: they are all related to the correlation function of the lowest component of the multiplet. For our purposes we need to find the form of the differential operator which relates the correlators $\langle \Psi^{(p_1)}\bar\Psi^{(p_2)}O^{(p_3)}O^{(p_4)}\rangle$ and $\langle \bar\Psi^{(p_1)}\Psi^{(p_2)}O^{(p_3)}O^{(p_4)}\rangle$ to $\langle O^{(p_1)} O^{(p_2)}O^{(p_3)}O^{(p_4)}\rangle$ \footnote{The relation between these four-point functions appears in \cite{Dolan:2001tt}, where the authors derived it by studying the supersymmetry transformation variations.}. In order to do that we will follow the same logic put forth in \cite{Belitsky:2014zha}. In the following we recall the main steps of the derivation. First, we strip off the part of four-point function which carries conformal weight
\begin{equation} \label{superward}
\mathcal{A}^{(p_1,p_2,p_3,p_4)}=\frac{y_{12}^{p_1+p_2}y_{34}^{p_3+p_4}}{x_{12}^{p_1+p_2}x_{34}^{p_3+p_4}}\left(\frac{x_{24} y_{14}}{x_{14} y_{24}} \right)^{p_{12}} \left(\frac{x_{14} y_{13}}{x_{13} y_{14}} \right)^{p_{34}} \hat{\mathcal{A}}\(x_i,y_i,\lambda_i,\pi_i\)\,,
\end{equation}
and introduce the function $\hat{\mathcal{A}}\(x_i,y_i,\lambda_i,\pi_i\)$ which should be annihilated by the supercharges $Q^b_{\alpha}$, $Q_{\alpha b'}$, $S^{\dot{\alpha} b}$ and $S^{\dot{\alpha}}_{b'}$ as well as $Q_{a \dot{\beta}}$, $Q^{a' \dot{\beta}}$, $S^{a' \beta}$ and $S_a^{\beta}$  to ensure its invariance under superconformal transformations. Following \cite{Belitsky:2014zha}, it is enough to impose  annihilation by half of the supercharges and the unique form for $\hat{\mathcal{A}}\(x_i,y_i,\lambda_i,\pi_i\)$ is given by
\begin{equation}
\hat{\mathcal{A}}\(x_i,y_i,\lambda_i,\pi_i\)=Q^4 Q'^4 S^4 S'^4 \lambda_1^4  \lambda_2^4  \lambda_3^4  \lambda_4^4 B(x_i,y_i)\,,
\end{equation}
where 
\begin{align}
Q^4&=\frac{1}{12} Q_a^\alpha Q_{\alpha}^b Q_{b}^{\beta}Q^a_{\beta}\,,& Q'^4&=\frac{1}{12} Q_{a'}^\alpha Q_{\alpha}^{b'} Q_{b'}^{\beta}Q^{a'}_{\beta}\,,& Q^{A}_{B}=\sum_{i=1}^4Q_{i,B}^A\,,\\
S^4&=\frac{1}{12}S_a^{\dot{\alpha}} S_{\dot{\alpha}}^b S_{b}^{\dot{\beta}}S^a_{\dot{\beta}} \,,& S'^4&=\frac{1}{12}S_{a'}^{\dot{\alpha}} S_{\dot{\alpha}}^{b'} S_{b'}^{\dot{\beta}}S^{a'}_{\dot{\beta}}\,,& S^{A}_{B}=\sum_{i=1}^4S_{i,B}^A\,,\\
 \lambda_i^4&=\lambda_{i a'}^{\alpha}\lambda_{i \alpha}^{b'} \lambda_{i b'}^{\beta} \lambda^{a'}_{i \beta}\, ,
\end{align}
and $B(x_i,y_i)$ is an arbitrary scalar function of the bosonic variables only. By equating the lowest component of \eqref{superward} to $A^{\{ p_i\}}(x_i,y_i)$, it is possible to find the relation between the function $B(x_i,y_i)$ and $A^{\{ p_i\}}(x_i,y_i)$. In order to simplify the computation we can fix a frame in which $x^{\alpha \dot{\alpha}}_3, y^{a a'}_3=0$ and $x_4^{\alpha \dot{\alpha}},y_4^ {a a'}\to \infty$. In this frame the function $B(x_i,y_i)$ reads
\begin{equation} \label{functionB}
B(x_i,y_i)=\frac{x_{12}^{p_1+p_2}x_{34}^{p_3+p_4}}{y_{12}^{p_1+p_2}y_{34}^{p_3+p_4}}\left(\frac{x_{24} y_{14}}{x_{14} y_{24}} \right)^{p_{21}} \left(\frac{x_{14} y_{13}}{x_{13} y_{14}} \right)^{p_{43}}\frac{A^{\{ p_i\}}(x_i,y_i)}{x_1^4 x_4^4 y_1^4 y_4^4 (z-\alpha)(z-\bar{\alpha})(\bar{z}-\alpha)(\bar{z}-\bar{\alpha})}\,.
\end{equation}
Notice that we do not take any limit in the prefactor.
We can  now apply this procedure to the correlation functions we are interested in and obtain
\begin{align} \label{fermionWard}
\langle \Psi^{(p_1)}\bar\Psi^{(p_2)}O^{(p_3)}O^{(p_4)}\rangle= \frac{y_{12}^{p_1+p_2}y_{34}^{p_3+p_4}}{x_{12}^{p_1+p_2}x_{34}^{p_3+p_4}}\left(\frac{x_{24} y_{14}}{x_{14} y_{24}} \right)^{p_{12}} \left(\frac{x_{14} y_{13}}{x_{13} y_{14}} \right)^{p_{34}} 
 \mathcal{M} \left(B(x_i,y_i)x_4^4 y_4^4\right)\,,
\end{align}
where $ \mathcal{M}$ is a differential operator which takes into account the projection on $\Psi$ and $\bar\Psi$ of the expression \eqref{superward}.  In the chosen frame it takes the form
\begin{equation}\label{Ward.M}
\mathcal{M}_{\alpha a',a\dot\alpha}=\partial_{x_{2}^{\beta \dot\alpha}}K_{\alpha a',a}^{\,\,\,\,\,\,\quad\beta}\,,
\end{equation}
with
\begin{align}\nonumber
K_{\alpha a',a}^{\,\,\,\,\,\,\quad\beta}&=x_{1,\alpha\dot\alpha}x_2^{\beta\dot\alpha}\left(y_{1,aa'}(x_{12}^2 y^2_2-x_2^2 y_{12}^2)+y_{12,aa'}(y_1^2 x_2^2-y_2^2 x_1^2)\right)\\\label{Ward.K}
&-x_{2,\alpha\dot\alpha}x_2^{\beta\dot\alpha}\left( y_2,aa'(x_1^2 y_{12}^2-x_{12}^2 y_1^2)+y_{21,aa'}(x_2^2 y^2_1-y^2_2 x^2_1)\right)\,.
\end{align}
One can also obtain an analogous relation for the correlator $\langle \bar\Psi^{(p_1)}\Psi^{(p_2)}O^{(p_3)}O^{(p_4)}\rangle$. In principle, we could reintroduce the dependence on $x_3, y_3$ and $x_4,y_4$ in the above expressions since we know the structure of free indices of $\mathcal{M}$ and the expressions for cross-ratios in the chosen frame:  $u= x_{12}/x_1^2$ and $v=x_2^2/x_1^2$. We will, however, postpone it and write in full generality the final expressions in the following section.

\section{Solution to the Casimir equation}\label{Sec:Solution}
\subsection{Deriving the differential equation}
In the previous sections we collected all ingredients necessary to find the solution to the Casimir equation \eqref{Casimireigenproblem}. In this section we use them in order to find an explicit form of the superconformal blocks. The strategy is to disentangle the lowest component of the superconformal block and write a differential equation for it. We accomplish it by projecting both sides of \eqref{Casimireigenproblem} to the lowest component, namely, evaluating it for $\lambda_i=0$ and $\pi_i=0$. On the right hand side of \eqref{Casimireigenproblem} it can be done immediately. The left hand side demands some work since the Casimir operator mixes various components of the superconformal block. 

Let us first understand which components of the superconformal block are relevant in our discussion. We can think about the superconformal block as a polynomial in Grassmann-odd variables $\lambda$ and $\pi$. Each component of the block is then multiplied by a Grassmann-odd polynomial of particular degree. Let us split the action of the two-particle Casimir in three pieces:
\begin{equation}
\mathcal{C}_{12}=\mathcal{C}_{\uparrow}+\mathcal{C}_0+\mathcal{C}_{\downarrow}\,,
\end{equation}
where operators $\mathcal{C}_{\uparrow}, \mathcal{C}_{\downarrow}$ and $\mathcal{C}_0$ increases, decreases and preserves the Grassmann-odd  degree, respectively. First, it is easy to see that we do not need the operator $\mathcal{C}_{\uparrow}$ at all, since it will produce a non-zero power of $\lambda$ or $\pi$, which eventually will vanish when projecting \eqref{Casimireigenproblem} on its lowest component. Second, the only relevant action of $\mathcal{C}_0$ is when we apply it on the lowest component of $\mathcal{G}_{\Delta,\ell,\mathcal{R}}^{\{ p_i\}}$. Finally, by inspecting the explicit form of the Casimir operator $\mathcal{C}_{\downarrow}$ we see that the only form it can take is $b(x,y)\partial_{\lambda_i}\partial_{\pi_j}$ for $i\neq j$ with some function $b(x,y)$\footnote{In general we expect also to find contributions with $i=j$. However, these cancel in the final form of the Casimir.}. Since the two-particle Casimir $\mathcal{C}_{12}$ acts only on the first two particles there are only two relevant contributions coming from $\mathcal{C}_{\downarrow}$, namely,
\begin{equation}
\mathcal{C}_{\downarrow}\sim b(x,y)\partial_{\lambda_1}\partial_{\pi_2}+\bar{b}(x,y)\partial_{\lambda_2}\partial_{\pi_1}\,,
\end{equation}
This explains why we need only $G^{\{p_i\}}_{\Delta,\ell,\mathcal{R}}$, $Q^{\{p_i\}}_{\Delta,\ell,\mathcal{R}}$ and $\bar{Q}^{\{p_i\}}_{\Delta,\ell,\mathcal{R}}$ in \eqref{superblock.expansion}.
To summarize, when projecting the Casimir eigenproblem \eqref{Casimireigenproblem} on its lowest component we end up with
\begin{equation}
\mathcal{C}_0 \, G^{\{p_i\}}_{\Delta,\ell,\mathcal{R}}+b(x,y)\,Q^{\{p_i\}}_{\Delta,\ell,\mathcal{R}}+\bar{b}(x,y)\,\bar{Q}^{\{p_i\}}_{\Delta,\ell,\mathcal{R}}=\kappa_{\Delta,\ell,\mathcal{R}}\,G^{\{p_i\}}_{\Delta,\ell,\mathcal{R}}\,,
\end{equation}
where $G^{\{p_i\}}_{\Delta,\ell,\mathcal{R}}$, $Q^{\{p_i\}}_{\Delta,\ell,\mathcal{R}}$ and $\bar{Q}^{\{p_i\}}_{\Delta,\ell,\mathcal{R}}$ are defined in \eqref{superblock.expansion}.

The degree-preserving part of the Casimir $\mathcal{C}_0$ can further be written as
\begin{equation}\label{bosonicblocksplit}
\mathcal{C}_0=\mathcal{C}_x-\mathcal{C}_y\,,
\end{equation}
where $\mathcal{C}_x$ (resp.~$\mathcal{C}_y$) is a differential operator depending only on variables $x$ (resp.~$y$) and equals the conformal symmetry quadratic Casimir (resp.~R-symmetry Casimir). The latter was constructed in e.~g.~\cite{Dolan:2003hv} and its action on $G^{\{ p_i\}}_{\Delta,\ell,\mathcal{R}}$ reads
\begin{align}\nonumber
\mathcal{C}_x \,G_{\Delta,\ell,\mathcal{R}}^{\{ p_i\}}&=\mathcal{C}_x \,g_{\Delta,\ell,\mathcal{R}}^{\{ p_i\}}+(p_{12}-p_{34})\left( (1+u-v)\left(u\frac{\partial}{\partial u}+v\frac{\partial}{\partial v}\right)-(1-u-v)\frac{\partial}{\partial v}\right)g_{\Delta,\ell,\mathcal{R}}^{\{ p_i\}}\\\label{Casimir.inhom}
&+\sfrac{1}{2}p_{12}\,p_{34}\,(1+u-v)g_{\Delta,\ell,\mathcal{R}}^{\{ p_i\}}\,,
\end{align}
with the action on any function of cross-ratios given by
\begin{align}\nonumber
\frac{1}{2}\mathcal{C}_x\, g_{\Delta,\ell,\mathcal{R}}^{\{ p_i\}}&=-\left((1-v)^2-u(1+v)\right) \frac{\partial}{\partial v}v\frac{\partial}{\partial v} g_{\Delta,\ell,\mathcal{R}}^{\{ p_i\}}-(1-u+v)u\frac{\partial}{\partial u}u\frac{\partial}{\partial u}g_{\Delta,\ell,\mathcal{R}}^{\{ p_i\}}\\\label{Casimir.hom}
&+2(1+u-v)u\, v\frac{\partial^2}{\partial u\partial v}g_{\Delta,\ell,\mathcal{R}}^{\{ p_i\}}+4u\frac{\partial}{\partial u}g_{\Delta,\ell,\mathcal{R}}^{\{ p_i\}}\,.
\end{align}
To get expressions for $\mathcal{C}_y$ we replace in \eqref{Casimir.inhom} and \eqref{Casimir.hom}: $u\to\sigma$, $v\to\tau$, $p_{12}\to-p_{12}$ and $p_{34}\to -p_{34}$.
The final step is to use the solution of superconformal Ward identities \eqref{Ward.M} together with \eqref{Ward.K} and combine it with the action of Casimir $\mathcal{C}_{\downarrow}$. We find
\begin{equation}\label{Cas.fer}
b(x,y)\,q^{\{p_i\}}_{\Delta,\ell,\mathcal{R}}+\bar{b}(x,y)\,\bar{q}^{\{p_i\}}_{\Delta,\ell,\mathcal{R}}=\mathcal{C}_{fer}\,g^{\{p_i\}}_{\Delta,\ell,\mathcal{R}}\,,
\end{equation}
where lower case functions $q^{\{p_i\}}_{\Delta,\ell,\mathcal{R}}$ and $\bar{q}^{\{p_i\}}_{\Delta,\ell,\mathcal{R}}$ are defined in analogy with \eqref{capitalG}. The explicit form of the operator on the right hand side of \eqref{Cas.fer} is 
\begin{align}\nonumber
\mathcal{C}_{fer}&=\frac{4 z(z-1)  (z (\alpha +\bar\alpha)-2 \alpha  \bar\alpha)}{(z-\alpha ) (z-\bar\alpha)}\frac{\partial}{\partial z}+\frac{4 \bar z(\bar z-1)  (\bar z (\alpha +\bar\alpha)-2 \alpha  \bar\alpha)}{(\bar z-\alpha ) (\bar z-\bar\alpha)}\frac{\partial}{\partial \bar z}\\\label{Casimir.fer}
&-\frac{4 \alpha(\alpha-1)  (\alpha (z +\bar z)-2 z  \bar z)}{(\alpha-z ) (\alpha-\bar z)}\frac{\partial}{\partial \alpha}-\frac{4 \bar \alpha(\bar\alpha-1)  (\bar\alpha (z +\bar z)-2 z  \bar z)}{(\bar\alpha-z ) (\bar\alpha-\bar z)}\frac{\partial}{\partial \bar\alpha}\,,
\end{align}
where the subscript $fer$ refers to the fact that this contribution originates from fermionic degrees of freedom of $\mathcal{N}=4$ SCFT. We use variables $z, \bar z,\alpha$ and $\bar\alpha$ to write \eqref{Casimir.fer} since it takes a simpler form compared to the original cross-ratios. 

Summarizing, we end up with the following equation for the lowest component of the superconformal block
\begin{equation}\label{final.equation}
\mathcal{C}_0 \, g^{\{p_i\}}_{\Delta,\ell,\mathcal{R}}+\mathcal{C}_{fer}\,g^{\{p_i\}}_{\Delta,\ell,\mathcal{R}}=\kappa_{\Delta,\ell,\mathcal{R}}\,g^{\{p_i\}}_{\Delta,\ell,\mathcal{R}}\,,
\end{equation}
where $\mathcal{C}_0$ is given in \eqref{bosonicblocksplit}, \eqref{Casimir.inhom} and \eqref{Casimir.hom}, $\mathcal{C}_{fer}$ is given in \eqref{Casimir.fer} and $\kappa_{\Delta,\ell,\mathcal{R}}$ in \eqref{Casimir.eigen}. We solve this equation in the following subsection. Finally, let us mention a simple yet important observation: from the explicit expressions for the Casimir \eqref{Casimir.inhom} and  \eqref{Casimir.fer} the symmetry between $u \leftrightarrow \sigma$, $v \leftrightarrow \tau$, or equivalently between $z \leftrightarrow \alpha$, $\bar{z} \leftrightarrow \bar{\alpha}$, is manifest. It turns out that the solutions to the Casimir eigenproblem, namely superconformal blocks, have the same symmetry and the only difference between space-time and R-symmetry parts are boundary conditions. More precisely, solutions to \eqref{final.equation} are polynomials in variables $y$ but more complicated functions of $x$, not even meromorphic. 

\subsection{Solutions to the Casimir equation} \label{Sec:solutionToCasimir}
From the study of superconformal Ward identities we know that there are two possible structures which can result in our study of conformal blocks. One is associated with long representations exchanges in the intermediate channel
\begin{equation}\label{block.long}
g^{\{p_i\}}_{\Delta,\ell,\mathcal{R}_L}\sim (z-\alpha)(z-\bar\alpha)(\bar z-\alpha)(\bar z-\bar\alpha)F(u,v,\sigma,\tau)\,,
\end{equation}
while the other is related to exchange of protected operators
\begin{align}\nonumber
g^{\{p_i\}}_{\Delta,\ell,\mathcal{R}_S}\sim&\left(\frac{\sigma}{u}\right)^{\sfrac{p_{34}}{2}} \left( -\frac{ (z-\alpha ) (z-\bar\alpha) (\bar z-\alpha ) f(z,\alpha )}{\alpha\,  z\, (\alpha -\bar\alpha) (z-\bar z)}+\frac{ (z-\alpha )
   (z-\bar\alpha) (\bar z-\bar\alpha) f(z,\bar\alpha)}{\bar\alpha\, z\, (\alpha -\bar\alpha) (z-\bar z)}\right. \\  \label{block.short}
   & \left. +\frac{ (z-\alpha )
   (\bar z-\alpha ) (\bar z-\bar\alpha) f(\bar z,\alpha )}{\alpha  \,\bar z\, (\alpha -\bar \alpha) (z-\bar z)}-\frac{ (z-\bar\alpha)
   (\bar z-\alpha ) (\bar z-\bar\alpha) f(\bar z,\bar\alpha)}{\bar \alpha\, \bar z\, (\alpha -\bar\alpha) (z-\bar z)}\right)\,.
\end{align}

\subsubsection{Long representations}
In order to solve \eqref{final.equation} for long representations we plug in the form \eqref{block.long} and notice that we can solve the derived equation for $F(u,v,\sigma,\tau)$ using the separation of variables technique. A similar equation has been obtained in the study of standard conformal blocks \cite{Dolan:2003hv} and we can closely follow here their derivation of solutions. We write
\begin{equation}\label{functionF}
F(u,v,\sigma,\tau)=H^{\{p_i\}}_{\Delta,\ell}(u,v)Y^{\{p_i\}}_{nm}(\sigma,\tau)\,.
\end{equation} 
Then one can find that the solution, up to an undetermined overall constant, is given by
\begin{align}
H^{\{ p_i\}}_{\Delta,\ell}(u,v)=(z\bar z)^{\sfrac{\Delta-\ell}{2}}g_{\Delta+4,\ell}(z,\bar z)\,,
\end{align}
where the function $g$ is the conformal block for non-supersymmetric CFTs and is given by
\begin{equation}
g_{\Delta,\ell}(z,\bar z)=\frac{1}{z-\bar z}\left[\left(\frac{-z}{2}\right)^\ell z\, k_{\Delta+\ell}(z)k_{\Delta-\ell-2}(\bar z)-\left(\frac{-\bar z}{2}\right)^\ell \bar z\, k_{\Delta+\ell}(\bar z)k_{\Delta-\ell-2}(z)  \right]\,,
\end{equation}
with 
\begin{equation}
k_{\beta}(x)={}_{2}F_{1}\left(\frac{\beta-p_{12}}{2},\frac{\beta+p_{34}}{2},\beta,x\right)\,.
\end{equation}
The R-symmetry part is given by
\begin{equation}
Y^{\{ p_i\}}_{nm}(\sigma,\tau)=\frac{P_{n+1}^{(a,b)}(y)P_{m}^{(a,b)}(\bar y)-P_{m}^{(a,b)}(y)P_{n+1}^{(a,b)}(\bar y)}{y-\bar y}\sigma^{\sfrac{p_{34}}{2}-2}\,,
\end{equation}
where
\begin{equation}\label{Jacobi.coeff}
a=\sfrac{p_{12}-p_{34}}{2}\,,\qquad b=\sfrac{-p_{12}-p_{34}}{2}\,,\qquad y=\sfrac{2}{\alpha}-1\,,\qquad \bar y=\sfrac{2}{\bar \alpha}-1\,,
\end{equation}
and $P^{(a,b)}_n(y)$ is the Jacobi polynomial.

It is easy to check that plugging this solution back in \eqref{Casimireigenproblem} we find 
\begin{equation}
\kappa_{\Delta,\ell,\mathcal{R}}=(\Delta+4)\Delta+\ell(\ell+2)-2(m(m+1)+n(n+3))+2p_{34}(n+m+2)+p_{34}^2\,,
\end{equation}
which is the eigenvalue of Casimir operator for long representation with dimension $\Delta$, spin $\ell$ and R-symmetry representation $\mathcal{R}=[n-m,2m-p_{34},n-m]$, as expected.

\subsubsection{Short representations}
For short and semi-short contributions we take \eqref{block.short} and plug it to the equation \eqref{final.equation}. In order to find the solution to this equation we focus on the leading expansion around $\bar z\sim \bar\alpha$, which will allow us to find the function $f( z, \alpha)$. Once again we can use the separation of variables
\begin{equation}
f(z,\alpha)=f_{\lambda}( z)P_{\mu}(\alpha)\,,
\end{equation}
and we find, up to an unfixed normalization, the following solution
\begin{equation}
f_{\lambda}( z)=z^{\lambda+1+\sfrac{p_{34}}{2}} \, _{2}F_{1}(\lambda+1-\sfrac{p_{12}}{2},\lambda+1+\sfrac{p_{34}}{2},2\lambda+2,z)\,,
\end{equation}
and
\begin{equation}
P_{\mu}(\alpha)=P^{(a,b)}_\mu(y)\,,
\end{equation}
where $P^{(a,b)}_\mu(y)$ is again the Jacobi polynomial and $a,b,y$ are given in \eqref{Jacobi.coeff}. 
It is  easy to check that when plugging this solution back in \eqref{final.equation} we get the following eigenvalue of Casimir operator
\begin{equation}
\kappa_{\Delta,\ell,\mathcal{R}}=2 (\lambda + 1) \lambda  - 2 (\mu +1-\sfrac{p_{34}}{2}) (\mu-\sfrac{p_{34}}{2}) \,.
\end{equation} 
It can be compared with the eigenvalue we expect for short and semi-short contributions. For short supermultiplets described in \eqref{short.dim} we get
\begin{equation}
\kappa_{2n,0,[n-m,m-\sfrac{p_{34}}{2},n-m]}=2(n-\sfrac{p_{34}}{2}+1)(n-\sfrac{p_{34}}{2})-2(m-\sfrac{p_{34}}{2}+1)(m-\sfrac{p_{34}}{2})\,,
\end{equation}
which corresponds to 
\begin{equation}
\lambda=n-\sfrac{p_{34}}{2}\,,\qquad \mu =m\,.
\end{equation}
For semi-short representations in \eqref{semishort.dim} the eigenvalue is 
\begin{equation}
\kappa_{2n+\ell+2,\ell,[n-m,m-\sfrac{p_{34}}{2},n-m]}=2(n+l+2-\sfrac{p_{34}}{2}+1)(n+l+2-\sfrac{p_{34}}{2})-2(m-\sfrac{p_{34}}{2}+1)(m-\sfrac{p_{34}}{2})\,,
\end{equation}
which agrees with our solution for
\begin{equation}
\lambda=n+l+2-\sfrac{p_{34}}{2}\,,\qquad \mu =m\,.
\end{equation}
A particular case of operator belonging to short representation is the identity operator. In this case the contribution is trivially given by the constant $k$ in \eqref{wardlowest} which comes from the contribution
\begin{equation}
f_{-1}(z)P_{0}(\alpha)=const\,.
\end{equation}

\section{Conclusions and outlook}
In this paper we have studied superconformal blocks of operators belonging to the half-BPS supermultiplet of $\mathcal{N}$=4 SCFT. We use the fact that superconformal blocks, as their conformal counterparts, are eigenfunctions of the two-particle quadratic Casimir operator of the superconformal algebra. We explicitly construct the two-particle quadratic super-Casimir operator. The action of this operator on the four-point function of operators in the half-BPS multiplets leads to a differential equation whose solutions give the form of the superconformal blocks. The way in which we computed superconformal blocks for the full supermultiplet is reminiscent of what was done for four-point functions. Also in that case, it is enough to know the four-point function of the lowest dimensional component of the supermultiplet to recover the four-point function of any other component of the supermultiplet by acting with a suitable differential operator \cite{Korchemsky:2015ssa}. 

To be more precise, having found the superconformal blocks one can expand the four-point function in superconformal partial waves as
\begin{equation} \label{superCPWA}
\mathcal{A}^{\{ p_i\}}=\frac{y_{12}^{p_1+p_2}y_{34}^{p_3+p_4}}{x_{12}^{p_1+p_2}x_{34}^{p_3+p_4}}\left(\frac{x_{24} y_{14}}{x_{14} y_{24}} \right)^{p_{12}} \left(\frac{x_{14} y_{13}}{x_{13} y_{14}} \right)^{p_{34}}\sum_{\Delta,\ell,\mathcal{R}}c_{\Delta,\ell,\mathcal{R}}^{p_1,p_2}c_{\Delta,\ell,\mathcal{R}}^{p_3,p_4} \,\mathfrak{g}^{\{p_i\}}_{\Delta,\ell,\mathcal{R}}(x_i,y_i,\lambda_i,\pi_i)\,,
\end{equation} 
where the sum runs over the quantum numbers (dimension, spin and $SU(4)_R$ representation labels) of the superconformal primaries appearing in the OPE of $O^{(p_1)} \times O^{(p_2)}$. Here the superconformal block is completely fixed by its lowest component and the superconformal Ward identity to be
\begin{equation}
\mathfrak{g}^{\{p_i\}}\(x_i,y_i,\lambda_i,\pi_i\)=Q^4 Q'^4 S^4 S'^4 \lambda_1^4  \lambda_2^4  \lambda_3^4  \lambda_4^4 \,g^{\{p_i\}}(u,v,\sigma,\tau)\,.
\end{equation}
 We would like to stress a difference in terminology compared to the results already present in the literature. In \cite{Dolan:2001tt} the notion of superconformal block was used to denote the function which appears in the decomposition of the lowest component of four-point function \eqref{4point}, namely \eqref{functionF}. In this paper we define as superconformal block the full eigenfunction of the supercasimir operator in such a way that it is possible to write down a decomposition of the form \eqref{superCPWA}, in full analogy with the conformal case. Then superconformal blocks are functions defined on the full analytic superspace and as such allow for instance to decompose in partial waves any component of the four-point function \eqref{4point}. 

 As we discussed already, for the superconformal blocks of the four-point function lowest component, the expression \eqref{superCPWA} reduces to two distinctive contributions
\begin{align} \nonumber
 A^{\{p_i\}}(x_i,y_i)&=\frac{y_{12}^{p_1+p_2}y_{34}^{p_3+p_4}}{x_{12}^{p_1+p_2}x_{34}^{p_3+p_4}}\left(\frac{x_{24} y_{14}}{x_{14} y_{24}} \right)^{p_{12}} \left(\frac{x_{14} y_{13}}{x_{13} y_{14}} \right)^{p_{34}} \left(\sum_{\Delta,\ell,\mathcal{R}_s}c_{\Delta,\ell,\mathcal{R}_S}^{p_1,p_2}c_{\Delta,\ell,\mathcal{R}_S}^{p_3,p_4} \,g^{\{p_i\}}_{\Delta,\ell,\mathcal{R}_S}(z,\bar{z}, \alpha, \bar{\alpha}) \right.\\\label{superCPWAsl}
&\left. +\sum_{\Delta,\ell,\mathcal{R}_L}c_{\Delta,\ell,\mathcal{R}_L}^{p_1,p_2}c_{\Delta,\ell,\mathcal{R}_L}^{p_3,p_4} \,g^{\{p_i\}}_{\Delta,\ell,\mathcal{R}_L}(z,\bar{z}, \alpha, \bar{\alpha})\right),
\end{align} 
where the subscript $S$ and $L$ denotes the contributions corresponding to the presence in the OPE of superconformal primaries belonging to short/semi-short and long representations, respectively. Notice that superconformal primaries which belong to long representations transform in smaller number of $SU(4)_R$ representations, as it is evident from the OPE decomposition \eqref{OPE}. Obviously superconformal descendants can transform in any representation consistent with the OPE. 

Since the three-point functions and the dimensions of half-BPS and quarter-BPS are protected from quantum correction, there is a subset of the sum in \eqref{superCPWAsl} which can be resumed to give an explicit, calculable function which depends only on variables $z$, $\bar{z}$, $\alpha$ and $\bar{\alpha}$ and possibly on $N$, rank of the gauge group. This is a non-trivial fact and it made possible to obtain a vast variety of results in the context of the numerical bootstrap \cite{Beem:2013qxa,Alday:2014qfa,Alday:2014tsa,Alday:2013opa}.

There are several ways in which our results can be useful:
\begin{itemize}
\item Having the explicit form of the superconformal blocks it is possible to use numerical bootstrap techniques to obtain further bounds on dimensions and OPE coefficients of superconformal primaries appearing in the OPE of any two operators belonging to the half-BPS multiplet of $\mathcal{N}$=4 SCFT. Moreover, due to the effectiveness and the success of the study of mixed correlators \cite{Kos:2014bka,Kos:2015mba} our results can be a starting point to perform such analysis also for four-dimensional $\mathcal{N}$=4 SYM. 
\item Recently the bootstrap equations have been studied also analytically \cite{Alday:2013cwa,Komargodski:2012ek,Fitzpatrick:2012yx,Alday:2015eya,Alday:2015ota,Kaviraj:2015xsa,Kaviraj:2015cxa}. In many of these studies, the form of the conformal blocks is crucial! Therefore one may hope to extract more analytic information for dimensions or OPE coefficients from the superconformal bootstrap equations. 
\item Although in our construction we have not referred to any  perturbative methods, one can extract perturbative CFT data using our superconformal blocks. The starting point would be to compute perturbatively four-point correlators of half-BPS states of the form we discuss in this paper. Then, using the superconformal partial wave decomposition, it is straightforward to extract the anomalous dimensions and the OPE coefficients, in a perturbative expansion, of the superconformal primaries appearing in the OPE of two external operators. Since we give the partial wave expansion for the full supermultiplets, it is possible using our results to compute these observables for more general classes of operators. Eventually, this may be of interest also in complementing results and techniques obtained using the powerful integrability methods in $\mathcal{N}$=4 SYM.
\item Our construction is based on the structure of the maximal superconformal symmetry algebra in four dimensions. It is, however, straightforward to generalize our formalism to theories with less amount of supersymmetry, as for example $\mathcal{N}$=2 conformal theories in four dimensions. Moreover, our study can be generalized to superconformal field theories living in different space-time dimensions, for instance three or six. Also for these cases it should be in principle possible to apply the same reasoning, provided that we found the solution to the superconformal Ward identities.
\item Lastly and more importantly, we believe that the approach used in this paper can be seen as a preliminary step in studying the superconformal blocks associated to four-point functions of generic long operators in $\mathcal{N}$=4 SYM. In that case there are several obstacles to overcome, as for example understanding the form of three-point function of two long operators and any operator appearing in their OPE. Differently from the case that we studied, those three-point functions are in general not related for different components of the same supermultiplet. However, pursuing such analysis would be very important to understand several features of non-perturbative $\mathcal{N}$=4 SYM data.
\end{itemize}

\section*{Acknowledgements}
We are very grateful to Livia Ferro, Paul Richmond and especially to Fernando Alday for many insightful discussions and comments. AB would like to thank the organizers of the Simons Summer Workshop 2015 at SCGP for hospitality and support, and the participants for many valuable discussions. This work was supported by ERC STG grant 306260.

\appendix
\section{Conventions}\label{App:Conventions}
Throughout the paper we use two-component spinor notation for both, Minkowski and R-symmetry indices. We adopt the following conventions for lowering and raising indices:
\begin{align}
\xi^{\alpha}=\epsilon^{\alpha\beta}\xi_{\beta}\,,\qquad  \xi^{\dot\alpha}=\epsilon^{\dot\alpha\dot\beta}\xi_{\dot\beta}\,,\qquad \xi^{a}=\epsilon^{ab}\xi_{b}\,,\qquad \xi^{a'}=\epsilon^{a'b'}\xi_{b'}\,,
\end{align}
with Levi-Civita symbol normalized as
\begin{align}
\epsilon^{\alpha\beta}=\epsilon_{\alpha\beta}=-\epsilon^{\dot\alpha\dot\beta}=-\epsilon_{\dot\alpha\dot\beta}=\epsilon^{ab}=\epsilon_{ab}=-\epsilon^{a'b'}=-\epsilon_{a'b'}=1\,.
\end{align}
Then, all distances are given by
\begin{align}
x^2=\sfrac{1}{2}x^{\alpha\dot\beta}x_{\alpha\dot\beta}\,,\qquad\qquad y^2=\sfrac{1}{2}y^{ab'}y_{ab'}\,.
\end{align}

\section{Form of generators}\label{App:Generators}
In this appendix we give the explicit form of generators of $\mathfrak{psu}(2,2|4)$ superconformal algebra realized as differential operators acting on the analytic superspace. All indices run over two values.

\subsection{Conformal algebra}
The conformal subalgebra consist of two $\mathfrak{su}(2)$ rotations $L_{\alpha}^{\,\,\,\beta}$ and $\bar{L}^{\dot\alpha}_{\,\,\,\dot\beta}$, translations $P_{\alpha\dot\beta}$, boosts $K^{\dot\alpha\beta}$ and the dilatation $D$:
\begin{align}
L_{\alpha}^{\,\,\,\beta}&=x^{\beta\dot\beta}\frac{\partial}{\partial x^{\alpha \dot\beta}}+\lambda^{\beta\a'}\frac{\partial}{\partial \lambda^{\alpha a'}}\,,\\
P_{\alpha\dot\beta}&=\frac{\partial}{\partial x^{\alpha\dot\beta}}\,,\\
K^{\dot\alpha\beta}&=x^{\gamma\dot\alpha}x^{\beta\dot\gamma}\frac{\partial}{\partial x^{\gamma\dot\gamma}}+x^{\gamma\dot\alpha}\lambda^{\beta c'}\frac{\partial}{\partial \lambda^{\gamma c'}}+\pi^{c\dot\alpha}x^{\beta\dot\gamma}\frac{\partial}{\partial \pi^{c\dot\gamma}}+\pi^{c\dot\alpha}\lambda^{\beta c'}\frac{\partial}{\partial y^{cc'}}\,,\\
\bar{L}^{\dot\alpha}_{\,\,\,\dot\beta}&=x^{\beta\dot\alpha}\frac{\partial}{\partial x^{\beta\dot\beta}}+\pi^{a\dot\alpha}\frac{\partial}{\partial \pi^{a\dot\beta}}\,,\\
D&=x^{\alpha\dot\alpha}\frac{\partial}{\partial x^{\alpha\dot\alpha}}+\sfrac{1}{2}\left(\lambda^{\alpha a'}\frac{\partial}{\partial \lambda^{\alpha a'}}+\pi^{a\dot\alpha}\frac{\partial}{\partial \pi^{a\dot \alpha}}\right)\,.
\end{align}

\subsection{R-symmetry algebra}
The R-symmetry algebra is realized in the analytic superspace in a direct analogue with the Minkowski space:
\begin{align}
R_{a}^{\,\,\,b}&=y^{b a'}\frac{\partial}{\partial y^{a a'}}+\pi^{b\dot\alpha}\frac{\partial}{\partial \pi^{a\dot\alpha}}\,,\\
P'_{a a'}&=\frac{\partial}{\partial y^{a a'}}\,,\\
K'^{\,a' a}&=y^{ba'}y^{ab'}\frac{\partial}{\partial y^{bb'}}+y^{ba'}\pi^{a\dot\beta}\frac{\partial}{\partial \pi^{b\dot\beta}}+\lambda^{\beta a'}y^{a b'}\frac{\partial}{\partial \lambda^{\beta b'}}+\lambda^{\beta a'}\pi^{a\dot\beta}\frac{\partial}{\partial x^{\beta\dot\beta}}\,,\\
\bar{R}^{a'}_{\,\,\,b'}&=y^{aa'}\frac{\partial}{\partial x^{a b'}}+\lambda^{\alpha a'}\frac{\partial}{\partial \lambda^{\alpha b'}}\,,\\
D'&=y^{a a'}\frac{\partial}{\partial y^{ab'}}+\sfrac{1}{2}\left(\lambda^{\alpha a'}\frac{\partial}{\partial \lambda^{\alpha a'}}+\pi^{a\dot\alpha}\frac{\partial}{\partial \pi^{a\dot \alpha}}\right)\,.
\end{align}

\subsection{Supercharges}
Supercharges split in two families: supertranslations $Q$ and superboosts $S$:
\begin{align}
Q_{\alpha}^{\,\,\, b}&=\pi^{b\dot\alpha}\frac{\partial}{\partial x^{\alpha\dot\alpha}}+y^{b a'}\frac{\partial}{\partial \lambda^{\alpha a'}}\,,\\
Q_{\alpha b'}&=\frac{\partial}{\partial \lambda^{\alpha b'}}\,,\\
Q_{a\dot\beta}&=\frac{\partial}{\partial \pi^{a\dot\beta}}\,,\\
Q^{a'}_{\,\,\, \dot\beta}&=\lambda^{\alpha a'}\frac{\partial}{\partial x^{\alpha\dot\beta}}-y^{aa'}\frac{\partial}{\partial \pi^{a\dot\beta}}\,,\\
S^{\dot\alpha b}&=x^{\beta\dot\alpha}\pi^{b\dot\beta}\frac{\partial}{\partial x^{\beta\dot\beta}}+x^{\beta\dot\alpha}y^{b b'}\frac{\partial}{\partial \lambda^{\beta b'}}-\pi^{c\dot\alpha}\pi^{b\dot\beta}\frac{\partial}{\partial \pi^{c\dot\beta}}-\pi^{c\dot\alpha}y^{b b'}\frac{\partial}{\partial y^{cb'}}\,,\\
S^{\dot\alpha}_{\,\,\,b'}&=x^{\alpha\dot\alpha}\frac{\partial}{\partial \lambda^{\alpha b'}}-\pi^{a\dot\alpha}\frac{\partial}{\partial y^{ab'}}\,,\\
S^{\,\,\,\beta}_{\alpha}&=x^{\beta\dot\alpha}\frac{\partial}{\partial \pi^{a\dot\alpha}}+\lambda^{\beta a'}\frac{\partial}{\partial y^{a a'}}\,,\\
S^{a' \beta}&=y^{b a'}\lambda^{\beta b'}\frac{\partial}{\partial y^{b b'}}+y^{b a'}x^{\beta\dot\beta}\frac{\partial}{\partial \pi^{b\dot\beta}}-\lambda^{\gamma a'}\lambda^{\beta b'}\frac{\partial}{\partial \lambda^{\gamma b'}}-\lambda^{\gamma a'}x^{\beta\dot\beta}\frac{\partial}{\partial x^{\gamma\dot\beta}}\,.
\end{align}

\bibliographystyle{nb}
\bibliography{bibliography}

\begin{thebibliography}{10}
\ifx\href\asklfhas\newcommand{\href}[2]{#2}\fi
\ifx\arxivref\asklfhas\newcommand{\arxivref}[2]{\href{http://arxiv.org/abs/#1}{#2}}\fi
\ifx\doiref\asklfhas\newcommand{\doiref}[2]{\href{http://dx.doi.org/#1}{#2}}\fi
\raggedright
\small
\parskip 0pt

\bibitem{Ferrara:1973yt}
S.~Ferrara, A.~F.~Grillo and R.~Gatto,
\textit{``{Tensor representations of conformal algebra and conformally
  covariant operator product expansion}''},
\textsf{\doiref{10.1016/0003-4916(73)90446-6}{Annals~Phys.~76,~161~(1973)}}.

\bibitem{Ferrara:1973vz}
S.~Ferrara, A.~F.~Grillo, G.~Parisi and R.~Gatto,
\textit{``{Covariant expansion of the conformal four-point function}''},
\textsf{\doiref{10.1016/0550-3213(72)90587-1}{Nucl.~Phys.~B49,~77~(1972)}}.

\bibitem{Polyakov:1974gs}
A.~M.~Polyakov,
\textit{``{Nonhamiltonian approach to conformal quantum field theory}''},
\textsf{Zh.~Eksp.~Teor.~Fiz.~66,~23~(1974)}.

\bibitem{Rattazzi:2008pe}
R.~Rattazzi, V.~S.~Rychkov, E.~Tonni and A.~Vichi,
\textit{``{Bounding scalar operator dimensions in 4D CFT}''},
\textsf{\doiref{10.1088/1126-6708/2008/12/031}{JHEP~0812,~031~(2008)}},
\texttt{\arxivref{0807.0004}{arxiv:0807.0004}}.

\bibitem{Poland:2010wg}
D.~Poland and D.~Simmons-Duffin,
\textit{``{Bounds on 4D Conformal and Superconformal Field Theories}''},
\textsf{\doiref{10.1007/JHEP05(2011)017}{JHEP~1105,~017~(2011)}},
\texttt{\arxivref{1009.2087}{arxiv:1009.2087}}.

\bibitem{ElShowk:2012hu}
S.~El-Showk and M.~F.~Paulos,
\textit{``{Bootstrapping Conformal Field Theories with the Extremal Functional
  Method}''},
\textsf{\doiref{10.1103/PhysRevLett.111.241601}{Phys.Rev.Lett.~111,~241601~(2013)}},
\texttt{\arxivref{1211.2810}{arxiv:1211.2810}}.

\bibitem{Caracciolo:2009bx}
F.~Caracciolo and V.~S.~Rychkov,
\textit{``{Rigorous Limits on the Interaction Strength in Quantum Field
  Theory}''},
\textsf{\doiref{10.1103/PhysRevD.81.085037}{Phys.~Rev.~D81,~085037~(2010)}},
\texttt{\arxivref{0912.2726}{arxiv:0912.2726}}.

\bibitem{ElShowk:2012ht}
S.~El-Showk, M.~F.~Paulos, D.~Poland, S.~Rychkov, D.~Simmons-Duffin and
  A.~Vichi,
\textit{``{Solving the 3D Ising Model with the Conformal Bootstrap}''},
\textsf{\doiref{10.1103/PhysRevD.86.025022}{Phys.~Rev.~D86,~025022~(2012)}},
\texttt{\arxivref{1203.6064}{arxiv:1203.6064}}.

\bibitem{El-Showk:2014dwa}
S.~El-Showk, M.~F.~Paulos, D.~Poland, S.~Rychkov, D.~Simmons-Duffin and
  A.~Vichi,
\textit{``{Solving the 3d Ising Model with the Conformal Bootstrap II.
  c-Minimization and Precise Critical Exponents}''},
\textsf{\doiref{10.1007/s10955-014-1042-7}{J.~Stat.~Phys.~157,~869~(2014)}},
\texttt{\arxivref{1403.4545}{arxiv:1403.4545}}.

\bibitem{Beem:2015aoa}
C.~Beem, M.~Lemos, L.~Rastelli and B.~C.~van~Rees,
\textit{``{The $(2,0)$ superconformal bootstrap}''},
\texttt{\arxivref{1507.05637}{arxiv:1507.05637}}.

\bibitem{Bobev:2015vsa}
N.~Bobev, S.~El-Showk, D.~Mazac and M.~F.~Paulos,
\textit{``{Bootstrapping the Three-Dimensional Supersymmetric Ising Model}''},
\textsf{\doiref{10.1103/PhysRevLett.115.051601}{Phys.~Rev.~Lett.~115,~051601~(2015)}},
\texttt{\arxivref{1502.04124}{arxiv:1502.04124}}.

\bibitem{Beem:2013qxa}
C.~Beem, L.~Rastelli and B.~C.~van~Rees,
\textit{``{The $\mathcal N=4$ Superconformal Bootstrap}''},
\textsf{\doiref{10.1103/PhysRevLett.111.071601}{Phys.Rev.Lett.~111,~071601~(2013)}},
\texttt{\arxivref{1304.1803}{arxiv:1304.1803}}.

\bibitem{Alday:2013opa}
L.~F.~Alday and A.~Bissi,
\textit{``{The superconformal bootstrap for structure constants}''},
\textsf{\doiref{10.1007/JHEP09(2014)144}{JHEP~1409,~144~(2014)}},
\texttt{\arxivref{1310.3757}{arxiv:1310.3757}}.

\bibitem{Alday:2014qfa}
L.~F.~Alday and A.~Bissi,
\textit{``{Generalized bootstrap equations for $ \mathcal{N}=4 $ SCFT}''},
\textsf{\doiref{10.1007/JHEP02(2015)101}{JHEP~1502,~101~(2015)}},
\texttt{\arxivref{1404.5864}{arxiv:1404.5864}}.

\bibitem{Chester:2014fya}
S.~M.~Chester, J.~Lee, S.~S.~Pufu and R.~Yacoby,
\textit{``{The $ \mathcal{N}=8 $ superconformal bootstrap in three
  dimensions}''},
\textsf{\doiref{10.1007/JHEP09(2014)143}{JHEP~1409,~143~(2014)}},
\texttt{\arxivref{1406.4814}{arxiv:1406.4814}}.

\bibitem{Bobev:2015jxa}
N.~Bobev, S.~El-Showk, D.~Mazac and M.~F.~Paulos,
\textit{``{Bootstrapping SCFTs with Four Supercharges}''},
\texttt{\arxivref{1503.02081}{arxiv:1503.02081}}.

\bibitem{Beem:2014zpa}
C.~Beem, M.~Lemos, P.~Liendo, L.~Rastelli and B.~C.~van~Rees,
\textit{``{The ${\mathcal N}=2$ superconformal bootstrap}''},
\texttt{\arxivref{1412.7541}{arxiv:1412.7541}}.

\bibitem{Khandker:2014mpa}
Z.~U.~Khandker, D.~Li, D.~Poland and D.~Simmons-Duffin,
\textit{``{$ \mathcal{N} $ = 1 superconformal blocks for general scalar
  operators}''},
\textsf{\doiref{10.1007/JHEP08(2014)049}{JHEP~1408,~049~(2014)}},
\texttt{\arxivref{1404.5300}{arxiv:1404.5300}}.

\bibitem{Fitzpatrick:2014oza}
A.~L.~Fitzpatrick, J.~Kaplan, Z.~U.~Khandker, D.~Li, D.~Poland and
  D.~Simmons-Duffin,
\textit{``{Covariant Approaches to Superconformal Blocks}''},
\textsf{\doiref{10.1007/JHEP08(2014)129}{JHEP~1408,~129~(2014)}},
\texttt{\arxivref{1402.1167}{arxiv:1402.1167}}.

\bibitem{Vichi:2011ux}
A.~Vichi,
\textit{``{Improved bounds for CFT's with global symmetries}''},
\textsf{\doiref{10.1007/JHEP01(2012)162}{JHEP~1201,~162~(2012)}},
\texttt{\arxivref{1106.4037}{arxiv:1106.4037}}.

\bibitem{Poland:2011ey}
D.~Poland, D.~Simmons-Duffin and A.~Vichi,
\textit{``{Carving Out the Space of 4D CFTs}''},
\textsf{\doiref{10.1007/JHEP05(2012)110}{JHEP~1205,~110~(2012)}},
\texttt{\arxivref{1109.5176}{arxiv:1109.5176}}.

\bibitem{Bashkirov:2013vya}
D.~Bashkirov,
\textit{``{Bootstrapping the $N=1$ SCFT in three dimensions}''},
\texttt{\arxivref{1310.8255}{arxiv:1310.8255}}.

\bibitem{Berkooz:2014yda}
M.~Berkooz, R.~Yacoby and A.~Zait,
\textit{``{Bounds on $\mathcal{N} = 1$ superconformal theories with global
  symmetries}''},
\textsf{\doiref{10.1007/JHEP01(2015)132,
  10.1007/JHEP08(2014)008}{JHEP~1408,~008~(2014)}},
\texttt{\arxivref{1402.6068}{arxiv:1402.6068}},
[Erratum: JHEP01,132(2015)].

\bibitem{Maldacena:1997re}
J.~M.~Maldacena,
\textit{``{The Large N limit of superconformal field theories and
  supergravity}''},
\textsf{\doiref{10.1023/A:1026654312961}{Int.~J.~Theor.~Phys.~38,~1113~(1999)}},
\texttt{\arxivref{hep-th/9711200}{hep-th/9711200}},
[Adv. Theor. Math. Phys.2,231(1998)].

\bibitem{Intriligator:1998ig}
K.~A.~Intriligator,
\textit{``{Bonus symmetries of N=4 superYang-Mills correlation functions via
  AdS duality}''},
\textsf{\doiref{10.1016/S0550-3213(99)00242-4}{Nucl.~Phys.~B551,~575~(1999)}},
\texttt{\arxivref{hep-th/9811047}{hep-th/9811047}}.

\bibitem{Intriligator:1999ff}
K.~A.~Intriligator and W.~Skiba,
\textit{``{Bonus symmetry and the operator product expansion of N=4
  SuperYang-Mills}''},
\textsf{\doiref{10.1016/S0550-3213(99)00430-7}{Nucl.~Phys.~B559,~165~(1999)}},
\texttt{\arxivref{hep-th/9905020}{hep-th/9905020}}.

\bibitem{Eden:1999gh}
B.~Eden, P.~S.~Howe and P.~C.~West,
\textit{``{Nilpotent invariants in N=4 SYM}''},
\textsf{\doiref{10.1016/S0370-2693(99)00705-4}{Phys.~Lett.~B463,~19~(1999)}},
\texttt{\arxivref{hep-th/9905085}{hep-th/9905085}}.

\bibitem{Petkou:1999fv}
A.~Petkou and K.~Skenderis,
\textit{``{A Nonrenormalization theorem for conformal anomalies}''},
\textsf{\doiref{10.1016/S0550-3213(99)00514-3}{Nucl.~Phys.~B561,~100~(1999)}},
\texttt{\arxivref{hep-th/9906030}{hep-th/9906030}}.

\bibitem{Howe:1999hz}
P.~S.~Howe, C.~Schubert, E.~Sokatchev and P.~C.~West,
\textit{``{Explicit construction of nilpotent covariants in N=4 SYM}''},
\textsf{\doiref{10.1016/S0550-3213(99)00768-3}{Nucl.~Phys.~B571,~71~(2000)}},
\texttt{\arxivref{hep-th/9910011}{hep-th/9910011}}.

\bibitem{Heslop:2001gp}
P.~J.~Heslop and P.~S.~Howe,
\textit{``{OPEs and three-point correlators of protected operators in N=4
  SYM}''},
\textsf{\doiref{10.1016/S0550-3213(02)00023-8}{Nucl.~Phys.~B626,~265~(2002)}},
\texttt{\arxivref{hep-th/0107212}{hep-th/0107212}}.

\bibitem{Eden:2001ec}
B.~Eden and E.~Sokatchev,
\textit{``{On the OPE of 1/2 BPS short operators in N=4 SCFT(4)}''},
\textsf{\doiref{10.1016/S0550-3213(01)00492-8}{Nucl.~Phys.~B618,~259~(2001)}},
\texttt{\arxivref{hep-th/0106249}{hep-th/0106249}}.

\bibitem{Dolan:2001tt}
F.~A.~Dolan and H.~Osborn,
\textit{``{Superconformal symmetry, correlation functions and the operator
  product expansion}''},
\textsf{\doiref{10.1016/S0550-3213(02)00096-2}{Nucl.~Phys.~B629,~3~(2002)}},
\texttt{\arxivref{hep-th/0112251}{hep-th/0112251}}.

\bibitem{Dolan:2004iy}
F.~A.~Dolan and H.~Osborn,
\textit{``{Conformal partial wave expansions for N=4 chiral four point
  functions}''},
\textsf{\doiref{10.1016/j.aop.2005.07.005}{Annals~Phys.~321,~581~(2006)}},
\texttt{\arxivref{hep-th/0412335}{hep-th/0412335}}.

\bibitem{Dolan:2004mu}
F.~A.~Dolan, L.~Gallot and E.~Sokatchev,
\textit{``{On four-point functions of 1/2-BPS operators in general
  dimensions}''},
\textsf{\doiref{10.1088/1126-6708/2004/09/056}{JHEP~0409,~056~(2004)}},
\texttt{\arxivref{hep-th/0405180}{hep-th/0405180}}.

\bibitem{Nirschl:2004pa}
M.~Nirschl and H.~Osborn,
\textit{``{Superconformal Ward identities and their solution}''},
\textsf{\doiref{10.1016/j.nuclphysb.2005.01.013}{Nucl.~Phys.~B711,~409~(2005)}},
\texttt{\arxivref{hep-th/0407060}{hep-th/0407060}}.

\bibitem{Dolan:2000ut}
F.~A.~Dolan and H.~Osborn,
\textit{``{Conformal four point functions and the operator product
  expansion}''},
\textsf{\doiref{10.1016/S0550-3213(01)00013-X}{Nucl.~Phys.~B599,~459~(2001)}},
\texttt{\arxivref{hep-th/0011040}{hep-th/0011040}}.

\bibitem{Howe:1995md}
P.~S.~Howe and G.~G.~Hartwell,
\textit{``{A Superspace survey}''},
\textsf{\doiref{10.1088/0264-9381/12/8/005}{Class.~Quant.~Grav.~12,~1823~(1995)}}.

\bibitem{Hartwell:1994rp}
G.~G.~Hartwell and P.~S.~Howe,
\textit{``{(N, p, q) harmonic superspace}''},
\textsf{\doiref{10.1142/S0217751X95001820}{Int.~J.~Mod.~Phys.~A10,~3901~(1995)}},
\texttt{\arxivref{hep-th/9412147}{hep-th/9412147}}.

\bibitem{Lee:1998bxa}
S.~Lee, S.~Minwalla, M.~Rangamani and N.~Seiberg,
\textit{``{Three point functions of chiral operators in D = 4, N=4 SYM at large
  N}''},
\textsf{Adv.~Theor.~Math.~Phys.~2,~697~(1998)},
\texttt{\arxivref{hep-th/9806074}{hep-th/9806074}}.

\bibitem{D'Hoker:1998tz}
E.~D'Hoker, D.~Z.~Freedman and W.~Skiba,
\textit{``{Field theory tests for correlators in the AdS / CFT
  correspondence}''},
\textsf{\doiref{10.1103/PhysRevD.59.045008}{Phys.~Rev.~D59,~045008~(1999)}},
\texttt{\arxivref{hep-th/9807098}{hep-th/9807098}}.

\bibitem{Howe:1998zi}
P.~S.~Howe, E.~Sokatchev and P.~C.~West,
\textit{``{Three point functions in N=4 Yang-Mills}''},
\textsf{\doiref{10.1016/S0370-2693(98)01431-2}{Phys.~Lett.~B444,~341~(1998)}},
\texttt{\arxivref{hep-th/9808162}{hep-th/9808162}}.

\bibitem{Penati:1999ba}
S.~Penati, A.~Santambrogio and D.~Zanon,
\textit{``{Two point functions of chiral operators in N=4 SYM at order
  g**4}''},
\textsf{\doiref{10.1088/1126-6708/1999/12/006}{JHEP~9912,~006~(1999)}},
\texttt{\arxivref{hep-th/9910197}{hep-th/9910197}}.

\bibitem{Penati:2000zv}
S.~Penati, A.~Santambrogio and D.~Zanon,
\textit{``{More on correlators and contact terms in N=4 SYM at order g**4}''},
\textsf{\doiref{10.1016/S0550-3213(00)00633-7}{Nucl.~Phys.~B593,~651~(2001)}},
\texttt{\arxivref{hep-th/0005223}{hep-th/0005223}}.

\bibitem{Korchemsky:2015ssa}
G.~P.~Korchemsky and E.~Sokatchev,
\textit{``{Four-Point Correlation Function of Stress-Energy Tensors in
  ${\mathcal N} = 4$ Superconformal Theories}''},
\texttt{\arxivref{1504.07904}{arxiv:1504.07904}}.

\bibitem{Belitsky:2014zha}
A.~V.~Belitsky, S.~Hohenegger, G.~P.~Korchemsky and E.~Sokatchev,
\textit{``{N=4 superconformal Ward identities for correlation functions}''},
\texttt{\arxivref{1409.2502}{arxiv:1409.2502}}.

\bibitem{PROP:PROP2190350705}
V.~K.~Dobrev and V.~B.~Petkova,
\textit{``Group-Theoretical Approach to Extended Conformal Supersymmetry:
  Function Space Realizations and Invariant Differential Operators''},
\textsf{\doiref{10.1002/prop.2190350705}{Fortschritte~der~Physik/Progress~of~Physics~35,~537~(1987)}}.

\bibitem{Beisert:2004ry}
N.~Beisert,
\textit{``{The Dilatation operator of N=4 super Yang-Mills theory and
  integrability}''},
\textsf{\doiref{10.1016/j.physrep.2004.09.007}{Phys.~Rept.~405,~1~(2004)}},
\texttt{\arxivref{hep-th/0407277}{hep-th/0407277}}.

\bibitem{Dolan:2002zh}
F.~A.~Dolan and H.~Osborn,
\textit{``{On short and semi-short representations for four-dimensional
  superconformal symmetry}''},
\textsf{\doiref{10.1016/S0003-4916(03)00074-5}{Annals~Phys.~307,~41~(2003)}},
\texttt{\arxivref{hep-th/0209056}{hep-th/0209056}}.

\bibitem{DOBREV1985127}
V.~Dobrev and V.~Petkova,
\textit{``All positive energy unitary irreducible representations of extended
  conformal supersymmetry''},
\textsf{\doiref{http://dx.doi.org/10.1016/0370-2693(85)91073-1}{Physics~Letters~B~162,~127
  ~(1985)}}.

\bibitem{Dolan:2003hv}
F.~A.~Dolan and H.~Osborn,
\textit{``{Conformal partial waves and the operator product expansion}''},
\textsf{\doiref{10.1016/j.nuclphysb.2003.11.016}{Nucl.~Phys.~B678,~491~(2004)}},
\texttt{\arxivref{hep-th/0309180}{hep-th/0309180}}.

\bibitem{Alday:2014tsa}
L.~F.~Alday, A.~Bissi and T.~Lukowski,
\textit{``{Lessons from crossing symmetry at large N}''},
\textsf{\doiref{10.1007/JHEP06(2015)074}{JHEP~1506,~074~(2015)}},
\texttt{\arxivref{1410.4717}{arxiv:1410.4717}}.

\bibitem{Kos:2014bka}
F.~Kos, D.~Poland and D.~Simmons-Duffin,
\textit{``{Bootstrapping Mixed Correlators in the 3D Ising Model}''},
\textsf{\doiref{10.1007/JHEP11(2014)109}{JHEP~1411,~109~(2014)}},
\texttt{\arxivref{1406.4858}{arxiv:1406.4858}}.

\bibitem{Kos:2015mba}
F.~Kos, D.~Poland, D.~Simmons-Duffin and A.~Vichi,
\textit{``{Bootstrapping the O(N) Archipelago}''},
\texttt{\arxivref{1504.07997}{arxiv:1504.07997}}.

\bibitem{Alday:2013cwa}
L.~F.~Alday and A.~Bissi,
\textit{``{Higher-spin correlators}''},
\textsf{\doiref{10.1007/JHEP10(2013)202}{JHEP~1310,~202~(2013)}},
\texttt{\arxivref{1305.4604}{arxiv:1305.4604}}.

\bibitem{Komargodski:2012ek}
Z.~Komargodski and A.~Zhiboedov,
\textit{``{Convexity and Liberation at Large Spin}''},
\textsf{\doiref{10.1007/JHEP11(2013)140}{JHEP~1311,~140~(2013)}},
\texttt{\arxivref{1212.4103}{arxiv:1212.4103}}.

\bibitem{Fitzpatrick:2012yx}
A.~L.~Fitzpatrick, J.~Kaplan, D.~Poland and D.~Simmons-Duffin,
\textit{``{The Analytic Bootstrap and AdS Superhorizon Locality}''},
\textsf{\doiref{10.1007/JHEP12(2013)004}{JHEP~1312,~004~(2013)}},
\texttt{\arxivref{1212.3616}{arxiv:1212.3616}}.

\bibitem{Alday:2015eya}
L.~F.~Alday, A.~Bissi and T.~Lukowski,
\textit{``{Large spin systematics in CFT}''},
\texttt{\arxivref{1502.07707}{arxiv:1502.07707}}.

\bibitem{Alday:2015ota}
L.~F.~Alday and A.~Zhiboedov,
\textit{``{Conformal Bootstrap With Slightly Broken Higher Spin Symmetry}''},
\texttt{\arxivref{1506.04659}{arxiv:1506.04659}}.

\bibitem{Kaviraj:2015xsa}
A.~Kaviraj, K.~Sen and A.~Sinha,
\textit{``{Universal anomalous dimensions at large spin and large twist}''},
\textsf{\doiref{10.1007/JHEP07(2015)026}{JHEP~1507,~026~(2015)}},
\texttt{\arxivref{1504.00772}{arxiv:1504.00772}}.

\bibitem{Kaviraj:2015cxa}
A.~Kaviraj, K.~Sen and A.~Sinha,
\textit{``{Analytic bootstrap at large spin}''},
\texttt{\arxivref{1502.01437}{arxiv:1502.01437}}.

\end{thebibliography}

\end{document}